\documentclass[aps, pra
, floatfix
, twocolumn
]{revtex4-1}

\usepackage{graphicx}
\usepackage{amsmath} 
\usepackage{amssymb}

\newcommand{\ket}[1]{\left| #1\right\rangle}
\newcommand{\bra}[1]{\left\langle #1\right|}
\newcommand{\braket}[2]{\left\langle #1 | #2 \right\rangle}

\begin{document}

\title{Optimal architecture for a non-deterministic noiseless linear
 amplifier} 
\author{N.~A.~McMahon} \author{A.~P.~Lund} \author{T.~C.~Ralph}
\affiliation{ Centre for Quantum Computation and Communications Technology,
School of Mathematics and Physics,\\
  University of Queensland, St Lucia Queensland 4072, Australia}

\begin{abstract}
  Non-deterministic quantum noiseless linear amplifiers are a new technology
  with interest in both fundamental understanding and new applications.  With a
  noiseless linear amplifier it is possible to perform tasks such as
  improving the performance of quantum key distribution and purifying lossy
  channels.  Previous designs for noiseless linear amplifiers involving linear
  optics and photon counting are non-optimal because they have a probability
  of success lower than the theoretical bound given by the theory of
  generalised quantum measurement.  This paper develops a theoretical model
  which reaches this limit.  We calculate the fidelity and probability of
  success of this new model for coherent states and Einstein-Podolsky-Rosen
  (EPR) entangled states.
\end{abstract}

\maketitle

A deterministic noiseless, phase insensitive, linear amplifier, as seen in
classical systems is unphysical in quantum theory~\cite{CAV82}. However it has
been demonstrated that an analogous probabilistic amplifier is approximately
physically realisable~\cite{ref:Ralph2008, ref:Xiang2010, ZAV11} and has a wide
variety of potential uses in quantum computing and communication technology
protocols. These protocols include error correction~\cite{ref:Ralph2011},
quantum key distribution~\cite{ref:Blandino2012}  and other protocols where
distillation of entanglement is desirable~\cite{ref:Xiang2010}.

In order to translate these systems to useful quantum technologies an
investigation into the optimal probabilities of success that can be achieved is
important. Low probabilities of success reduce the range of possible
experimental and commercial applications of these devices. Ralph and Lund
\cite{ref:Ralph2008} proposed a linear optics implementation of a heralded
noiseless linear amplifier which has been theoretically investigated
\cite{FIU09, GAG12, ref:Walk2013} and experimentally demonstrated with good
agreement in visibility and effective gain for small amplitudes $\alpha < 0.04$
and gains $|g|^{2}\leq 5$ \cite{ref:Xiang2010, ref:Ferreyrol2008, MIC12,OSO12,
KOS13}. The probability of success for low amplitude inputs $\alpha \ll 1$
using this design is $P = \frac{1}{g^{2}+1}$. The probability of success of
other linear optical designs are similar \cite{ZAV11, KIM12}. For higher
amplitudes, $\bar n$, the probability scales as $P \approx
\frac{1}{(g^{2}+1)^N}$ where $N \gg |\alpha|^2$.  The theoretical maximum
probability of success for a noiseless linear amplifier in the low photon
number regime is $P = \frac{1}{g^{2}}~$\cite{ref:Ralph2008} and is expected to
scale as $\frac{1}{g^{2N}}$. 

Our aim in this paper is to identify and analyse a physical model for noiseless
linear amplification which saturates this maximum probability of success. Our
approach is related to the idea that noiseless amplification can be implemented
via a weak measurement model~\cite{ref:Menzies2009}. The paper is arranged in
the following way. In the first section we will introduce a measurement model
for noiseless amplification. In section 2 we will translate this into a
physical model for the amplifier and particularly look at the low photon number
limit. The following two sections will analyse the performance of the amplifier
with respect to coherent state inputs and the distillation and purification of
Einstein, Podolsky, Rosen (EPR) entanglement (2-mode squeezing). In the final
section we will conclude.

\section{Noiseless amplification as a general measurement}

An ideal noiseless amplifier performs the operation $g^{a^\dagger
  a}$~\cite{ref:Ralph2008}, that is it takes an input state $\ket{\psi}$ to
$g^{a^\dagger a}\ket{\psi}$.  This operator takes the coherent state
$\ket{\alpha}$ to the coherent state $\ket{g\alpha}$ and is inherently not
unitary.  This suggests that a measurement process with post-selection on the
measurement outcomes is required to implement it.  The case we are most
interested in here is where $g > 1$.  In this situation the operator is
unbounded and can only be implemented perfectly over the entire Hilbert space
via a measurement process with probability zero.  In many experimental
situations the action of this operator on states with high occupation number
are not important as they have negligible amplitude.  Therefore this operator
is generally chosen to be truncated at some occupation number $N$, which will
be chosen depending on the desired performance of an experimental apparatus.
This truncation allows for non-zero probabilities of successfully implementing
the desired amplification transformation.  Lower values of $N$ will generally
result in higher probabilities of success at the cost of a lower fidelity of
operation when compared to the ideal operation.  In current experiments with
low energy inputs $N=1$ is sufficient to achieve high fidelity, and this very
simple case has non-trivial implications.

When constructing a measurement which implements the amplification, it
suffices to consider the case where there is only two outcomes, a success
outcome and a failure outcome.  When a success outcome is achieved the state
is transformed in the required way.  Measurement outcomes, which we will label
$i$ are represented by $S$ for success and $F$ for failure.  The action on the
input state due to each measurement result can be represented by the generally
non-unitary operator $\hat{M}_{i}$ called the measurement operator. The
probability of success for this measurement outcome when the measurement is
applied to the state $\ket{\psi}$ is given by
\begin{equation}
  P_{i} = \bra{\psi} \hat{M}_{i}^{\dagger} \hat{M}_{i} \ket{\psi}
\end{equation}
and the resultant output state having achieved the result $i$ is
\begin{equation}
  \ket{\psi^{'}_{i}} = \frac{\hat{M}_{i} \ket{\psi}}{\sqrt{P_i}}.
\end{equation}
To ensure that these operators define a probability measure the condition
\begin{equation}
  \label{eq:MeasurementModelRestriction}
  \hat{M}_S \hat{M}^\dagger_S + \hat{M}_F\hat{M}^\dagger_F = \hat{I}
\end{equation}
must be satisfied~\cite{ref:MikeandIke}. 

To implement the amplification we require $\hat{M}_S \propto g^{a^\dagger a}$.
To ensure \eqref{eq:MeasurementModelRestriction} holds over the entire Hilbert
space it would be necessary for $\hat{M}_S = 0 g^{\hat{a}^\dagger \hat{a}}$ as
the eigenvalues of $g^{\hat{n}}$ are unbounded for $g > 1$ and
$\hat{M}_F\hat{M}_F^\dagger$ must be a positive operator.  Now we can make the
truncation of this operator to achieve a non-zero probability.  We do this by
requiring the action on the first $N$ Fock states to be proportional to the
those same elements for the perfect amplification operator and leaving the
action on higher occupation number states arbitrary. In this case the success
measurement operator can be written as
\begin{equation}
  \hat{M}_S = \mathcal{N} \sum_{n = 0}^{N} g^{n}\ket{n}\bra{n} +
  \sum_{n = N+1}^{\infty} S_{n} \ket{n}\bra{n},
\end{equation}
where $S_{n}$ is a sequence of complex numbers with norm between zero and
one. This will then allow the operation to satisfy
\eqref{eq:MeasurementModelRestriction} with $\mathcal{N}$ playing the role of
the proportionality constant and will in general be non-zero.  The probability
of success for an arbitrary input state $\ket{\psi}$ is
\begin{widetext}
\begin{equation}
  \label{eq:Probability}
  P_S = \bra{\psi} M_{S}^{\dagger} M_{S} \ket{\psi}
  = \left|\mathcal{N}\right|^{2} 
  \sum_{n=0}^{N} g^{2n} \left| \braket{n}{\psi}\right|^{2} 
  + \sum_{n=N+1}^{\infty} \left| S_{n}\right|^{2} \left|\braket{n}{\psi}\right|^{2}.
\end{equation}
To ensure that $0 \leq P_S \leq 1$ for all possible input states $\mathcal{N}
\leq g^{-N}$.  Here we can see that any complex phase factor within each $S_n$
will not influence the probability of success.  The fidelity of the success
operation for pure state inputs is
\begin{equation}
  \label{eq:Fidelity}
  \mathcal{F}
  = \frac{\left|\bra{\psi} g^{a^\dagger a} M_S \ket{\psi}\right|^{2}}
  {\bra{\psi} M_{S}^{\dagger} M_{S} \ket{\psi}} 
  = P^{-1} \left| 
    \mathcal{N} \sum_{n=0}^{N} g^{2n} \left|\braket{n}{\psi}\right|^2 
    + \sum_{n=N+1}^{\infty} S_n g^n \left|\braket{n}{\psi}\right|^2\right|^2.
\end{equation}
\end{widetext}
Here the complex phase factors of the $S_n$ are important.  However if the
$S_n$ are not real than this can only act to reduce the fidelity.  Therefore,
to maximise the fidelity and probability over the widest set of states then
requires $\mathcal{N} = g^{-N}$ and $S_n = 1$.  This optimised measurement
operator is then
\begin{equation}
  \label{eq:SuccessMO}
  \hat{M}_{S} = g^{-N}\sum_{n = 0}^{N} g^{n}\ket{n}\bra{n} 
  + \sum_{n = N+1}^{\infty} \ket{n}\bra{n}
\end{equation}

\section{Measurement model for noiseless amplification}

\begin{figure}
\includegraphics{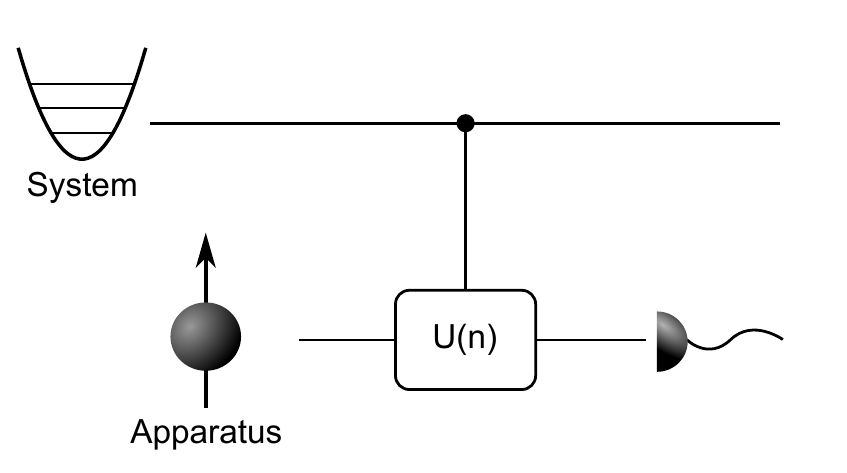}
\caption{A bosonic system (labeled ``System'') interacts with a two-level
  apparatus (labeled ``Apparatus'').  The apparatus is prepared into a $Z$
  axis spin eigenstate.  The interaction applies a conditional unitary
  rotation where the conditioning depends on the number of bosons in the
  input.  The apparatus is measured and if the spin has flipped then a success
  is heralded.}
\label{fig:model}
\end{figure}

We can construct a model for the generalised measurement described in
Eq.~\ref{eq:SuccessMO} by considering a measurement apparatus consisting of a
two level system which interacts with the bosonic input mode as shown in
Figure~\ref{fig:model}.
After the interaction the apparatus is measured using a projective measurement
scheme.  The apparatus orthonormal basis states represent success and failure
and will be written as $\ket{S}$ and $\ket{F}$ respectively.  This basis is
arbitrary, but the interaction will depend on the particular choice of basis.
We will assume that the apparatus is prepared in the $\ket{F}$ state before
the interaction.  The interaction is given by the unitary operator
\begin{equation}
\hat{U} = \hat{M}_{S} \otimes \ket{S} \bra{F} +
\hat{M}_{F} \otimes \ket{F} \bra{F} + 
\hat{B}_{1} \otimes \ket{F} \bra{S} + 
\hat{B}_{2} \otimes \ket{S} \bra{S}
\end{equation}
where $\hat{M}_{S}$ is the operator which will be applied to the system input
state when a success result is measured and $\hat{M}_{F}$ is the operator
applied to the system on measuring the failure result.  The particular form of
the operators $\hat{B}_{1,2}$ are not of concern as they are dependent on the
apparatus being initialised in the $\ket{S}$ state.  They are included to
include enough freedom to ensure that $\hat{U}$ remains unitary.  Using the
Kronecker product representation of the tensor product the unitarity
requirement can be written as
\begin{equation}
  \left( 
    \begin{matrix} 
      \hat{M}^{\dagger}_{F} & \hat{M}^{\dagger}_{S} \\
      \hat{B}^{\dagger}_{1} & \hat{B}^{\dagger}_{2} 
    \end{matrix}
  \right) 
  \left(
    \begin{matrix}
      \hat{M}_{F} & \hat{B}_{1} \\
      \hat{M}_{S} & \hat{B}_{2} 
    \end{matrix}
  \right) 
= \left( 
  \begin{matrix} 
    \hat{I} & 0 \\
    0 & \hat{I}
  \end{matrix}
\right)
\end{equation}
which can be rewritten as
\begin{eqnarray}
  \hat{M}_{F}^{\dagger}\hat{M}_{F} + \hat{M}_{S}^{\dagger}\hat{M}_{S} &=& \hat{I} \\
  \hat{M}_{F}^{\dagger} \hat{B}_{1} + \hat{M}_{S}^{\dagger} \hat{B}_{2} &=& 0 \\
  \hat{B}_{1}^{\dagger}\hat{B}_{1} + \hat{B}_{2}^{\dagger}\hat{B}_{2} &=& \hat{I}
\label{eq:Requirements}
\end{eqnarray}
Provided $\hat{M}_{S}$ and $\hat{M}_{F}$ define a set of measurement operators
(in particular the requirement in equation
\eqref{eq:MeasurementModelRestriction}) then the first and last equations are
always satisfied if $\hat{B}_{1} = \pm \hat{M}_{S}$ and $\hat{B}_{2} = \pm
\hat{M}_{F}$.  The second equation could never be satisfied had we swapped the
success and failure operators in this assignment.  If $\hat{M}_{S}$ and
$\hat{M}_{F}$ are Hermitian and commute, as is the case we are considering
here, then we can always satisfy the second equation by choosing $\hat{B}_{1} =
-\hat{M}_{S}$ and $\hat{B_{2}} = \hat{M}_{F}$.

Now we can substitute our success operator from equation~\eqref{eq:SuccessMO}
into this interaction unitary.  This unitary can then be rearranged to be
written as 
\begin{equation}
	\label{eq:AmpUnitary}
	\hat{U} = \sum_{n=0}^{\infty} \ket{n}\bra{n} \otimes \hat{R}_{n}
\end{equation}
where $\hat{R}_{n}$ is defined as
\begin{equation}
  \label{eq:Rotation}
  \hat{R}_{n} = 
  \left( 
    \begin{matrix}
      \sqrt{1-G_{n}^{2}} & G_{n} \\
      -G_{n} & \sqrt{1-G_{n}^{2}} 
    \end{matrix} 
  \right), 
\end{equation}
\begin{equation}
  G_{n} = \min(1,g^{(n-N)}).
\end{equation}
The operator $\hat{R}_{n}$ is a Pauli Y-rotation of $\theta =
2\arcsin\left(\min(1, g^{n-N})\right)$ on the heralding qubit which depends on
the number of bosons in the input mode.  This unitary can be generated by the
Hamiltonian
\begin{eqnarray}
  \label{eq:Hamiltonian}
  \hat{H} &=& \frac{\hbar}{\tau} \left[
    \sum_{n = 0}^{\infty} \arcsin(\min(1,g^{n-N})) \hat{Y} \otimes \ket{n}\bra{n}
  \right] \nonumber \\
  &=& \frac{\hbar}{\tau} \arcsin(\min(1,g^{\hat{a}^\dagger \hat{a}-N})) \otimes \hat{Y},
\end{eqnarray}
where $\tau$ is the interaction time which is chosen to ensure the apropriate
that the rotation parameter $\theta$ is implemented.

\subsection{Low photon number limit} 

In the limit of low amplitude inputs we can implement the amplifier with $N=1$.
The system can then be considered a qubit and the gate between the system and
the apparatus is locally equivalent to a standard controlled rotation.
To see this, we take the unitary from equation~\ref{eq:AmpUnitary}
\begin{equation}
	\hat{U}_{N=1} =	\ket{0}\bra{0} \otimes \left(
	\begin{matrix}
		\sqrt{1-1/g^2} & 1/g \\
		-1/g & \sqrt{1-1/g^2}
	\end{matrix}
	\right) +
	\ket{1}\bra{1} \otimes \left(
	\begin{matrix}
		0 & 1 \\
		-1 & 0
	\end{matrix}
	\right)
\end{equation}
and then decompose it into
\begin{equation}
	\hat{U}_{N=1} = -(X\otimes X) (I\otimes Z) C(R_y(\theta)) (X\otimes I)
\end{equation}
where $X$ and $Z$ are the standard Pauli matricies and $C(R_y(\theta))$ is a
controlled Pauli $Y$ rotation by $\theta$ and $\theta$ is as defined above with
$n=0$ and $N=1$.  Applying this unitary to states of the form $\ket{0} + \alpha
\ket{1}$ where $\alpha$ is small results in the probability of success for the
noiseless amplification of $\frac{1}{g^2}$.

\section{Coherent state inputs}

\begin{figure}
  \includegraphics{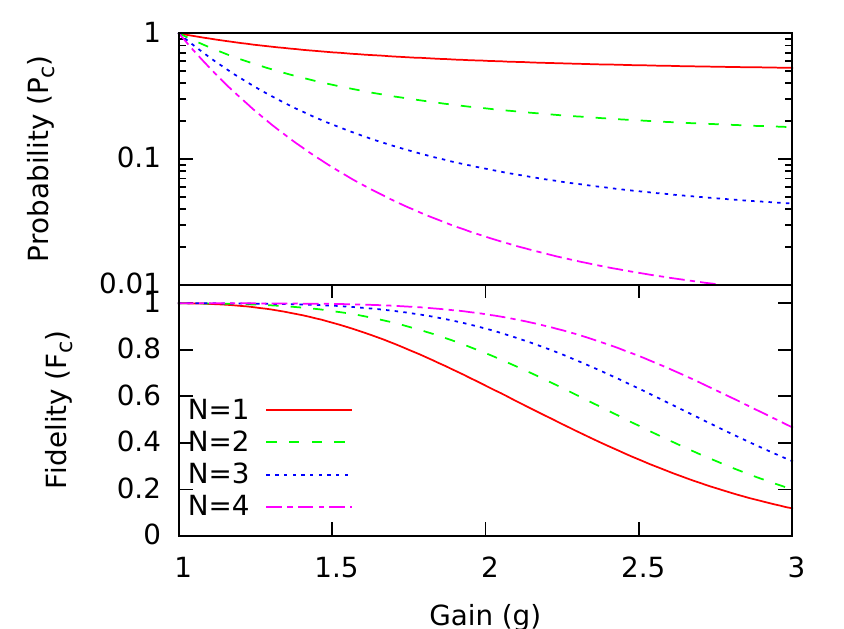}
  \caption{Probability of success and fidelity for an input coherent state
    with amplitude $\alpha=0.8$ for $N=1,2,3$ and $4$.  These curves are
    calculated from equations~\ref{eq:coherentprobability}
    and~\ref{eq:coherentfidelity}.}
  \label{fig:coherent0.8}
\end{figure}

We can now calculate the performance of this model for particular situations.
First we will calculate the action on coherent states.  Coherent states are an
ideal test of the amplification process as the expected output from the
amplification is easy to define.  The ideal amplification action on a coherent
state is
\begin{equation}
g^{a^\dagger a} \ket{\alpha} = e^{(g^2-1)|\alpha|^2/2}\ket{g\alpha}.
\end{equation}
This can then be used to calculate the probability of success and the fidelity
of our model amplifier for coherent state inputs denoted by $P_c$ and
$\mathcal{F}_c$ respectively,
\begin{widetext}
\begin{equation}
  \label{eq:ProbabilityCoherent}
  P_c = \bra{\alpha} M_S^{\dagger} M_S \ket{\alpha} 
  = e^{-\left|\alpha\right|^{2}} 
  \left[g^{-2N} \sum_{n = 0}^{N} g^{2n} \frac{\left|\alpha\right|^{2n}}{n!} + 
    \sum_{n = N+1}^{\infty} \frac{\left|\alpha\right|^{2n}}{n!}\right],
\end{equation}
\begin{equation}
  \label{eq:FidelityCoherent}
  \mathcal{F}_c
  = P_c^{-1} \left|\bra{g \alpha} M_{s} \ket{\alpha}\right|^{2}
  = P_c^{-1} e^{-(1+g^2)\left|\alpha\right|^{2}} 
  \left| g^{-N} \sum_{n=0}^{N} g^{2n} \frac{\left|\alpha\right|^{2n}}{n!} 
   + \sum_{n=N+1}^{\infty} g^{n} \frac{\left|\alpha\right|^{2n}}{n!}\right|^{2}.
\end{equation}
These expressions can be written in terms of incomplete gamma functions
\begin{equation}
  \label{eq:coherentprobability}
  P_c = P(N+1,|\alpha|^2) + g^{-2N} e^{(g^2-1)|\alpha|^2} Q(N+1, |g\alpha|^2),
\end{equation}
\begin{equation}
  \label{eq:coherentfidelity}
  F_c = P_c^{-1} e^{-(1+g^2)|\alpha|^2} \left| g^{-N} e^{|g\alpha|^2} P(N+1,|g\alpha|^2)
  + e^{g|\alpha|^2} Q(N+1, g|\alpha|^2) \right|^2
\end{equation}
\end{widetext}
where $Q(N,\lambda)$ is the regularlised incomplete gamma function defined as
\begin{equation}
 Q(N,\lambda)=\Gamma(N,\lambda)/\Gamma(N) 
\end{equation}
where $\Gamma(N,\lambda)$ is the incomplete gamma function, $\Gamma(N)$ is the
complete gamma function and $P(N,\lambda) = 1 - Q(N,\lambda)$~\cite{gammafunc}.
The appearance of the incomplete gamma functions here is expected as this
function is the cumulative distribution function for the Possionian
distribution which is the distribution that would result when measuring a
coherent state in the Fock basis.  In this form these equations can be rapidly
computed numerically for particular values of $g$, $\alpha$ and $N$.
Figure~\ref{fig:coherent0.8} shows $P_c$ and $F_c$ for $\alpha=0.8$ and
$N=1,2,3$ and $4$.
The probability drops away from $1$ for small gains and the rate at which this
occurs increases as $N$ increases.  The fidelity initially stays close to $1$
for small amplitudes but eventually drops and the gain at which this occurs
increases as $N$ increases.  Whilst these properties are evident in the
figure, they are general features given that $\alpha$ is fixed.

\begin{figure}
  \includegraphics{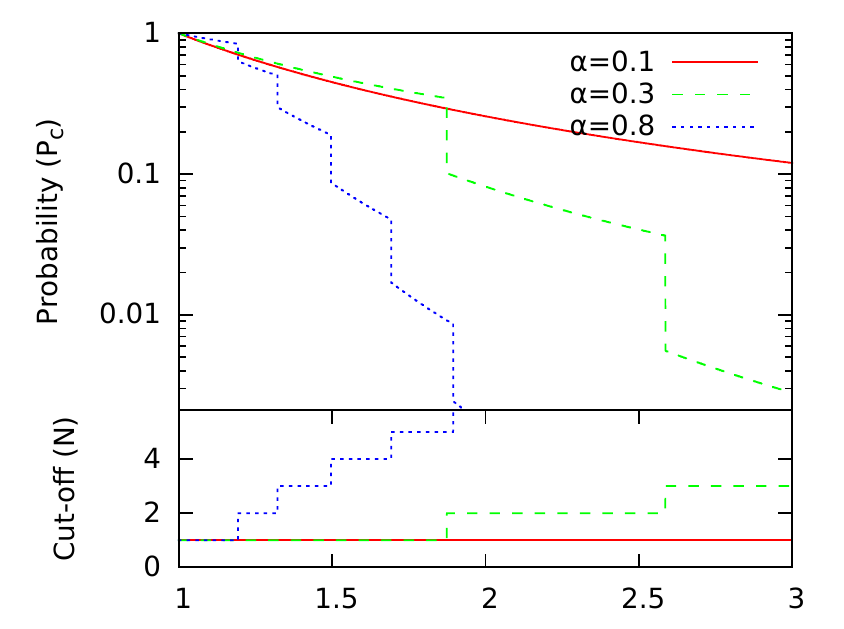}
  \caption{Probability of success for coherent state inputs with amplitude
    $\alpha=0.1,0.3,0.8$ for gains between $g=1$ and $4$.  Cut-off $N$ is
    chosen to ensure an output fidelity more than $0.99$.  Discontinuous jumps
    occur when the fidelity bound is reached and the value of $N$ is
    incremented.  The corresponding values for $N$ are shown in the lower
    plot.}
  \label{fig:GraphsCoherent}
\end{figure}

Low fidelity operation is not of great interest for building a device which
performs linear amplification.  Therefore we will set a bound on performance
that is deemed acceptable.  Quantitatively we will require a minimum fidelity
$\mathcal{F} \geq 0.99$.  The fidelity will increase towards $1$ as $N$
increases hence in any particular situation we can choose an $N$ to achieve
this fidelity requirement.  Figure~\ref{fig:GraphsCoherent} shows the effect
of enforcing this minimum acceptable fidelity.
The most notable effect that can be seen is the discontinuous jumps in the
probability of success.  A jump occurs when the cut-off $N$ is incremented to
enforce the minimum fidelity.  This means that the probability of success is
made up of pieces from the probabilities like what is shown in
figure~\ref{fig:coherent0.8} for $\alpha=0.8$.  Also of note, is that for low
amplitude inputs (here $\alpha=0.1$) then choosing $N=1$ provides an
acceptable reproduction of linear amplification over a wide range of gain
(here $1 \leq g \leq 3$).

\section{EPR state inputs}

An important application of this type of amplification is distilling
continuous variable entanglement~\cite{ref:Xiang2010,ref:Blandino2012}.  The
action of the amplifier is easiest to calculate for an ideal
Einstein-Podolsky-Rosen (EPR) state
\begin{equation}
\label{eq:EPRState}
\ket{EPR} = \sqrt{1-\chi^{2}} \sum_{n=0}^\infty \chi^n \ket{n, n},
\end{equation}
where the parameter $0\leq\chi<1$ is representative of the strength of the
continuous variable entanglement.  The ideal amplification of this state is
then
\begin{equation}
g^{a^\dagger a}\ket{EPR} \propto \sum_{n=0}^\infty (g \chi)^n \ket{n,n}.
\end{equation}
The action of the amplifier preserves the form of the EPR state but increases
the entanglement.  Note that this places an upper bound on $g$.  For if $g >
1/\chi$ then the coefficients in the summation diverge.  What this means is
that when an implementation chooses an $N$ cut-off, the output state does not
converge towards a particular state in the limit as $N \rightarrow \infty$.
This phenomenon will also found when applying ideal amplification to a
distribution of coherent states which forms a mixed state~\cite{ref:Walk2013}.

The EPR state can be generalised to include losses.  Here we will concentrate
on the case where only one of the EPR modes undergoes loss of amplitude
$\eta$.  The state from this is a three mode state
\begin{equation}
  \label{eq:EPRloss}
  \frac{\ket{EPR_l}}{\sqrt{1-\chi^{2}}} =  \sum_{n=0}^{\infty} \sum_{t=0}^{n} \chi^{n}
  \sqrt{\binom{n}{t} \eta^{t} \left(1-\eta\right)^{n-t}} \ket{n, t, n-t}
\end{equation}
where the third mode represents the loss mode which is assumed to be
inaccessible to any experiment.

\begin{figure}
\includegraphics[scale = 0.2]{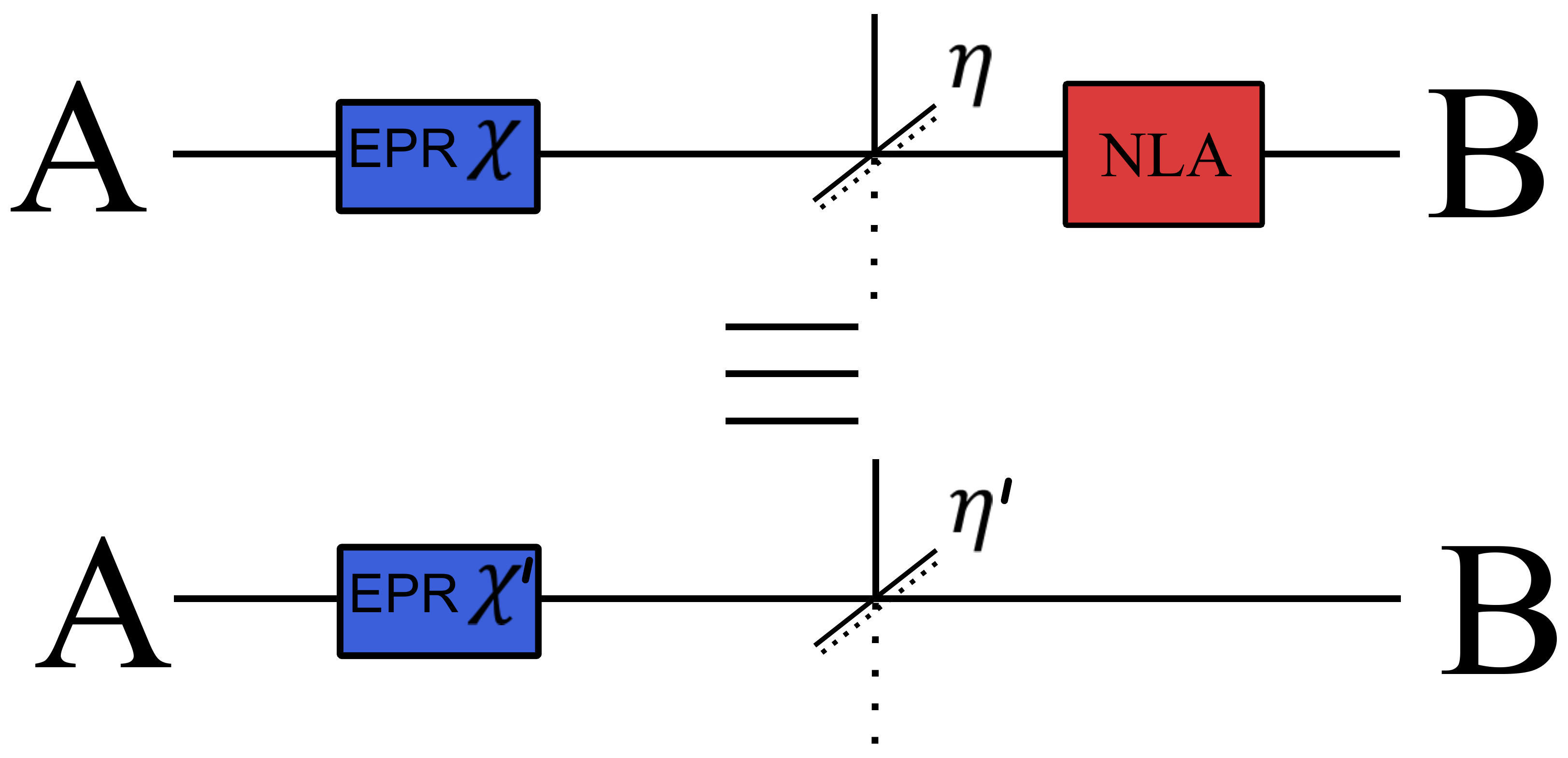}
\caption{\small{The state generated by an ideal noiseless linear amplifier on
    a single sided lossy EPR state is another single sided lossy EPR state but
    with different variables for the strength of the squeezing and loss.  The
    parameters of the state after the amplification $\chi^\prime$ and
    $\eta^\prime$ are related to the input state parameters $\chi$ and $\eta$
    and the gain of the amplification $g$
    (equations~\ref{eq:ChiRelation},~\ref{eq:EtaRelation}
    and~\ref{eq:AuxRelation}).}}
\label{fig:EPREquiv} 
\end{figure}
As in the case of the pure EPR state, the lossy EPR state under ideal
amplification is another lossy EPR state but with different parameters, see
figure~\ref{fig:EPREquiv}.  Applying the ideal amplification to the second
mode in equation~\ref{eq:EPRloss} introduces a $g^t$ into the coefficients.
Then equating this to another lossy EPR state characterised by squeezing
$\chi^\prime$ and transmission $\eta^\prime$ gives the relations
\begin{equation}
\chi^n g^t \sqrt{\eta}^t \left(\sqrt{1-\eta}\right)^{n-t} =
{\chi^\prime}^n \sqrt{\eta^\prime}^t \left(\sqrt{1-\eta^\prime}\right)^{n-t}
\end{equation}
which must hold true for all integers $n \geq 0$ and $0\leq t \leq n$.  Two
separate equations can be obtained from this,
\begin{equation}
\chi \sqrt{1-\eta} = \chi^\prime \sqrt{1-\eta^\prime}
\end{equation}
\begin{equation}
\chi g \sqrt{\eta} = \chi^\prime \sqrt{\eta^\prime},
\end{equation} 
which can be inverted to give
\begin{equation}
\label{eq:ChiRelation}
\chi^\prime = f \chi,
\end{equation}
\begin{equation}
\label{eq:EtaRelation}
\eta^\prime = \frac{g^2}{f^2} \eta,
\end{equation}
\begin{equation}
  \label{eq:AuxRelation}
  f = \sqrt{1-\eta + \eta g^2}.
\end{equation}
The possibility of non-convergence of the output state, just as seen for pure
EPR inputs, is present here as well.  Convergence will be achieved provided
$\chi^\prime < 1$.

We will consider $\eta$ to be a fixed value and choose $\chi^\prime$ to be a
fixed in the sense that some target squeezing strength is desired.  In this
way we can avoid choosing gains for which the output is not convergent.

We will focus here on the ability of the state to demonstrate the EPR
paradox~\cite{reid1998,ou1992}.  This is achieved by EPR criterion
$\varepsilon_{EPR} < 1$ where
\begin{equation}
\varepsilon_{EPR} = V_{B|A}^+ V_{B|A}^-
\end{equation}
and $V_{B|A}^\pm$ is the conditional variance of the $B$ mode on $A$ and the
superscript represents the quadrature in which the variance is calculated.
The conditional variance is defined as
\begin{equation}
V_{B|A}^\pm = \min_{0 \leq g \leq 1} \left< (X_B^\pm \mp g X_A^\pm)^2 \right>,
\end{equation}
and for the EPR state with one sided loss the optimisation gives~\cite{bernu} 
\begin{equation}
V^+_{B|A} = V^-_{B|A} = 1 - \frac{2 \chi^2 \eta}{1+\chi^2},
\end{equation}
and hence the EPR criterion in this case is
\begin{equation}
\label{eq:EPRCondition}
\varepsilon_{EPR} = \left(1-\frac{2\chi^2 \eta}{1+\chi^2}\right)^{2}.
\end{equation}
When the amplifier succeeds, both the effective squeezing and transmission are
greater then their initial counterparts.  The amplifier has a purifing action
on this state. This means that it is possible to reach a lower EPR
criterion then would be otherwise possible.
\begin{figure}
\includegraphics{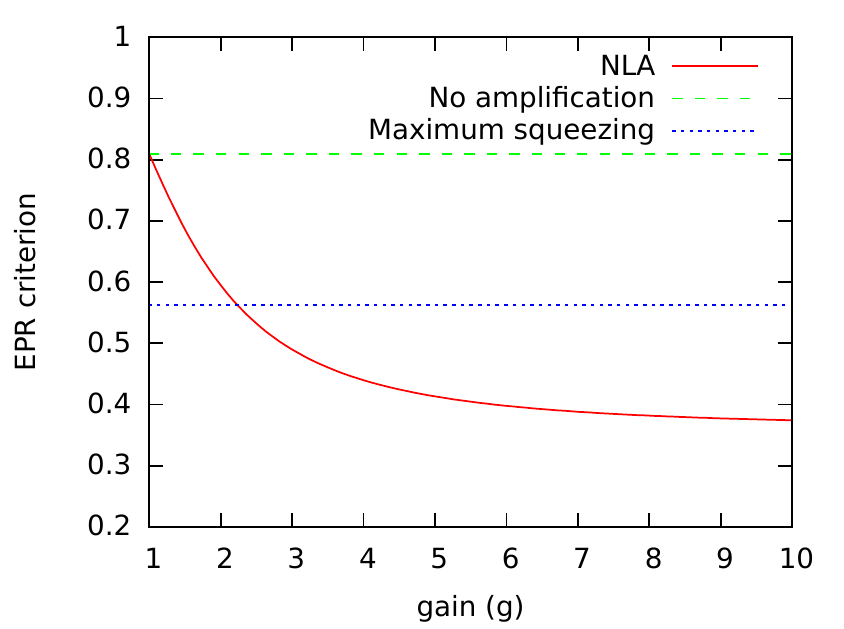}
\caption{The EPR criterion as a function of gain with a target output
  squeezing of $\chi^{'} = 0.5$ and an initial transmission of $\eta =
  0.25$. The red line indicates the initial EPR criterion (no amplification),
  while the green line represents the maximum EPR criterion that could be
  obtained by using a squeezed state with $\chi \rightarrow 1$ without making
  use of a noiseless amplifier.}
  \label{fig:EPRIdeal}
\end{figure}

Figure~\ref{fig:EPRIdeal} shows the EPR criterion for an output squeezing of
$\chi^{'} = 0.5$ with a channel transmission of $\eta = 0.25$. The lowest EPR
condition possible without amplification given the channel loss (i.e. $\chi
\rightarrow 1$) is achieved by amplifying the lossy EPR state when $g\approx
2.5$. 

The state conditional on achieving success is
\begin{widetext}
\begin{equation}
  \label{eq:EPRsuccess}
  M_s \ket{EPR_l} = \sum_{t=0}^\infty \sum_{n=t}^\infty 
    \mathrm{min}(g^{t-N},1) \chi^n 
    \sqrt{\binom{n}{t} \eta^t (1-\eta)^{n-t}} \ket{n,t,n-t}.
\end{equation}
The probability of success for our model amplifier on this type of input state
can be simply computed as $P_{EPR} = \bra{EPR}\hat{M}_{S}^{\dagger}
\hat{M}_{S} \ket{EPR}$ just as before
\begin{equation}
  \label{eq:ProbabilityEPR}
  P_{EPR} = (1-\chi^2)\left(
    \frac{g^{-2N}}{1-\chi^2(1+(g^2-1)\eta)} +
    \sum_{n=N+1}^\infty \chi^{2n}
    \sum_{t=N+1}^n \binom{n}{t} \left(1-g^{2(t-N)}\right) \eta^t (1-\eta)^{n-t}\
  \right).
\end{equation}
A sum can be removed from this equation by using the relationship
\begin{equation}
  \sum_{t=N+1}^n \binom{n}{t} a^t b^{n-t} =
(a+b)^n I_\frac{a}{a+b}(N+1,n-N),
\end{equation}
where $I_x(a,b)$ is the regularised incomplete beta function~\cite{betafunc},
giving \begin{equation}
    P_{EPR} = (1-\chi^2)\left(
      \frac{g^{-2N}}{1-{\chi^\prime}^2} +
      \sum_{n=N+1}^\infty  \left(
        \chi^{2n} I_\eta(N+1,n-N) 
        - g^{-2N} {\chi^\prime}^{2n} I_{\eta^\prime}(N+1,n-N)
      \right)
    \right). 
  \end{equation}
To compute fidelity is more difficult because when the loss mode is traced out
the resulting state is mixed.  We can calculate a lower bound on the fidelity
by computing the fidelity of the amplified state compared to the purified
lossy EPR state with squeezing $\chi^\prime$ and loss $\eta^\prime$,
i.e. $\mathcal{F}_{EPR} = P_{EPR}^{-1} \left|\bra{EPR^{'}}\hat{M}_{S}
  \ket{EPR}\right|^{2}$
\begin{equation}
  \frac{\sqrt{\mathcal{F}_{EPR} P_{EPR}}}{\sqrt{(1-\chi^2)(1-\chi^{\prime 2})}} =
  \frac{g^{-N}}{1-(g\sqrt{\eta\eta^\prime}+\sqrt{(1-\eta)(1-\eta^\prime)})\chi \chi^\prime}
  + \sum_{n=N+1}^\infty (\chi \chi^\prime)^n \sum_{t=N+1}^n
  \binom{n}{t} (1 - g^{t-N}) \sqrt{\eta\eta^\prime}^t
  \sqrt{(1-\eta)(1-\eta^\prime)}^{n-t}
\end{equation}
\begin{equation}
  \eta_1 = \sqrt{\eta\eta^\prime} + \sqrt{(1-\eta)(1-\eta^\prime)} 
  = \frac{1-\eta+g\eta}{f}
\end{equation}
\begin{equation}
  \eta_2 = g\sqrt{\eta\eta^\prime} + \sqrt{(1-\eta)(1-\eta^\prime)} = f
\end{equation}
\begin{equation}
  \frac{\sqrt{\mathcal{F}_{EPR} P_{EPR}}}{\sqrt{(1-\chi^2)(1-\chi^{\prime 2})}} =
  \frac{g^{-N}}{1-\eta_2 \chi \chi^\prime}
  + \sum_{n=N+1}^\infty (\chi \chi^\prime)^n \left( 
    \eta_1^n I_{\sqrt{\eta\eta^\prime}/\eta_1} (N+1,n-N)
    - g^{-N} \eta_2^n I_{g\sqrt{\eta\eta^\prime}/\eta_2} (N+1,n-N)
  \right)
\end{equation}
\end{widetext}
where $f$ is defined in Equation~\ref{eq:AuxRelation}.

\begin{figure*}
  \includegraphics{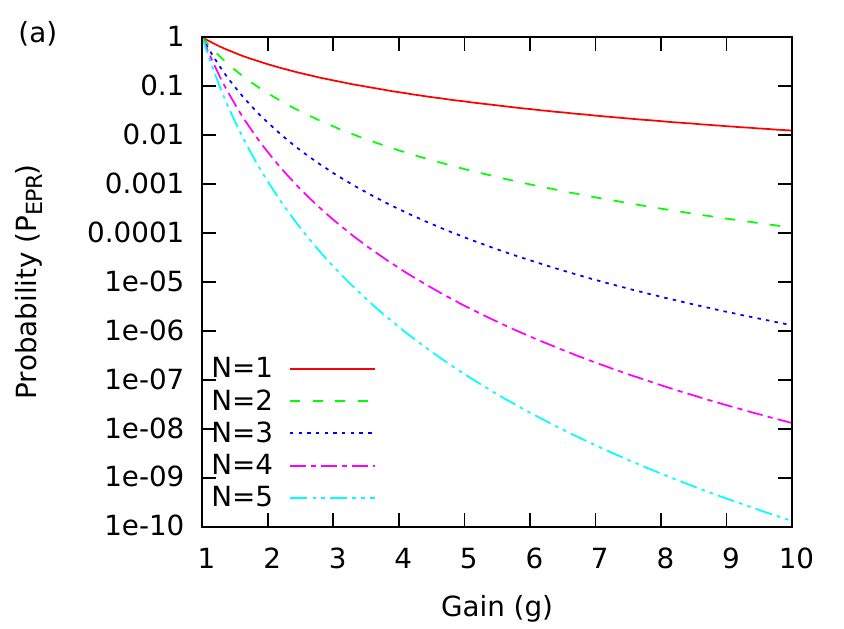}
  \includegraphics{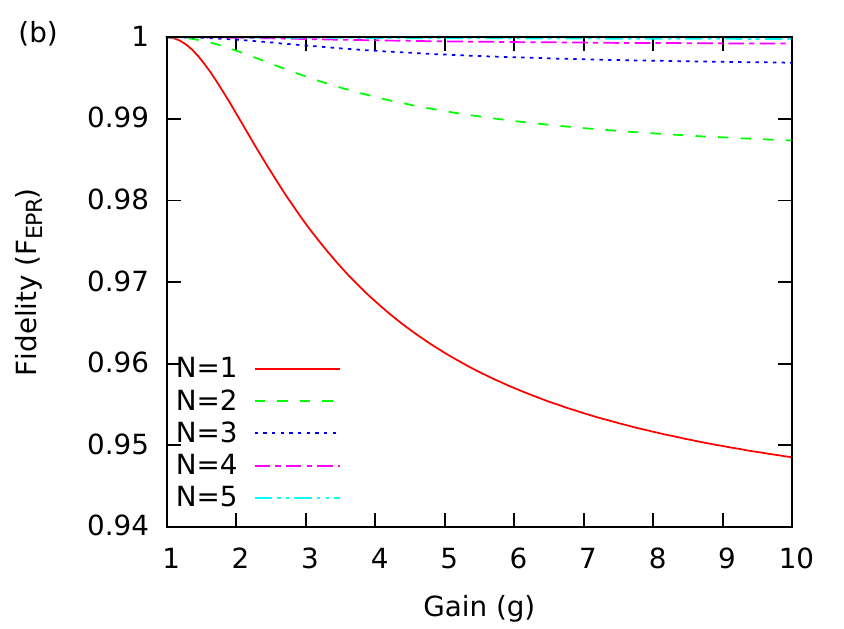}
  \caption{Probability and fidelity for the EPR state characterised by an
    effective squeezing of $\chi' = 0.5$ and a transmission of $\eta =
    0.3$ undergoing amplification with truncation numbers $N = 1$ to $N =
    5$.} 
  \label{fig:EPRRealNLA} 
\end{figure*}
The probability and fidelity for $N = 1$ to $5$ with $\chi^\prime = 0.5$ and
$\eta=0.3$ are shown in figure \ref{fig:EPRRealNLA}.  The probabilities drop
exponentially with gain, but the fidelity drops slowly.  This is because as
the gain increases a lower $\chi$ is used to ensure that $\chi^\prime$ stays
fixed.  The asymptotic behaviour of these functions as $g \rightarrow \infty$ 
is  
\begin{equation}
	P_{EPR} = g^{-2N} \left(
	\frac{1-{\chi^\prime}^{2N+2}}{1-{\chi^\prime}^2}
	\right) + O (g^{-2N-1}),
\end{equation}
\begin{equation}
	\mathcal{F}_{EPR} P_{EPR} = g^{-2N} 
	\frac{(1-{\chi^\prime}^{2N+2})^2}{1-{\chi^\prime}^2} + O(g^{-2N-1}).
\end{equation}
Hence we find that the fidelity asymptotically approaches a constant value
\begin{equation}
	\mathcal{F}_{EPR} = 1-{\chi^\prime}^{2N+2} + O(g^{-1}).
\end{equation}
The fidelity will always be $1$ at $g = 1$ and for larger $g$ then approaches
this constant value from above.  Therefore this number constitutes a lower
bound on the fidelity.

As was indicated before in the analysis for coherent state inputs, the low
fidelity operation is not usually of interest.  When designing an experiment
there is usually some minimum fidelity and probability of success that is
deemed acceptable.  The order of magnitude for these is dependant on the on the
experimental conditions.  We will now consider these factors to further analyse
the action of this model amplifier.

We can use this expression for the limiting case of fidelity to explicly
compute a maximum $N$ under restrictions in the fidelity and entanglement.   A
fidelity minimum is chosen $\mathcal{F}_{min} < 1$ and at all times the
performance of amplification must always be higher than this number.  Also, if
there is a maximum $\chi^\prime < 1$ for which amplifications cannot exceed
after successful amplification, then it must be true that 
\begin{equation}
	N \leq 
	\frac{\log{\left(1-\mathcal{F}_{min}\right)}}{2 \log{\chi^\prime}} - 1
\end{equation}
Note that this requirement is independent of the probability of success.

\begin{figure*}[p!]
  \includegraphics[type=pdf,ext=.pdf,read=.pdf]{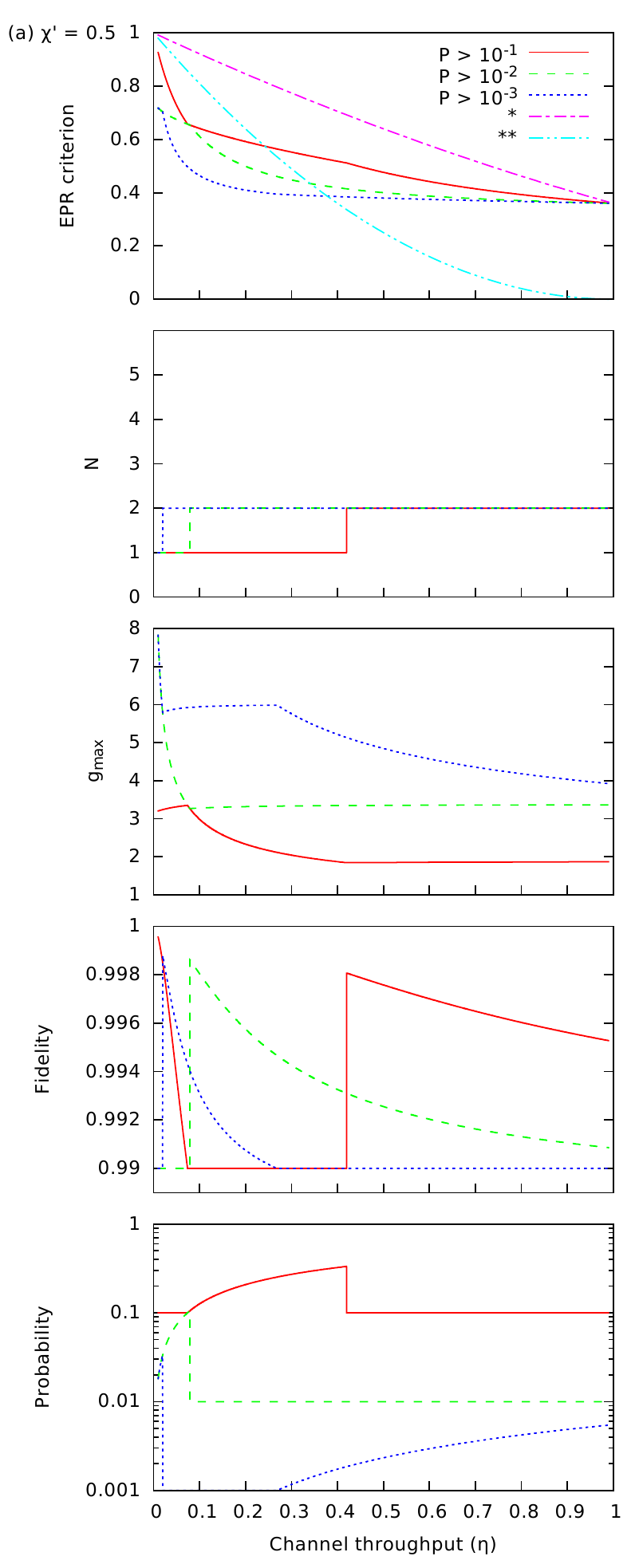}
  \includegraphics[type=pdf,ext=.pdf,read=.pdf]{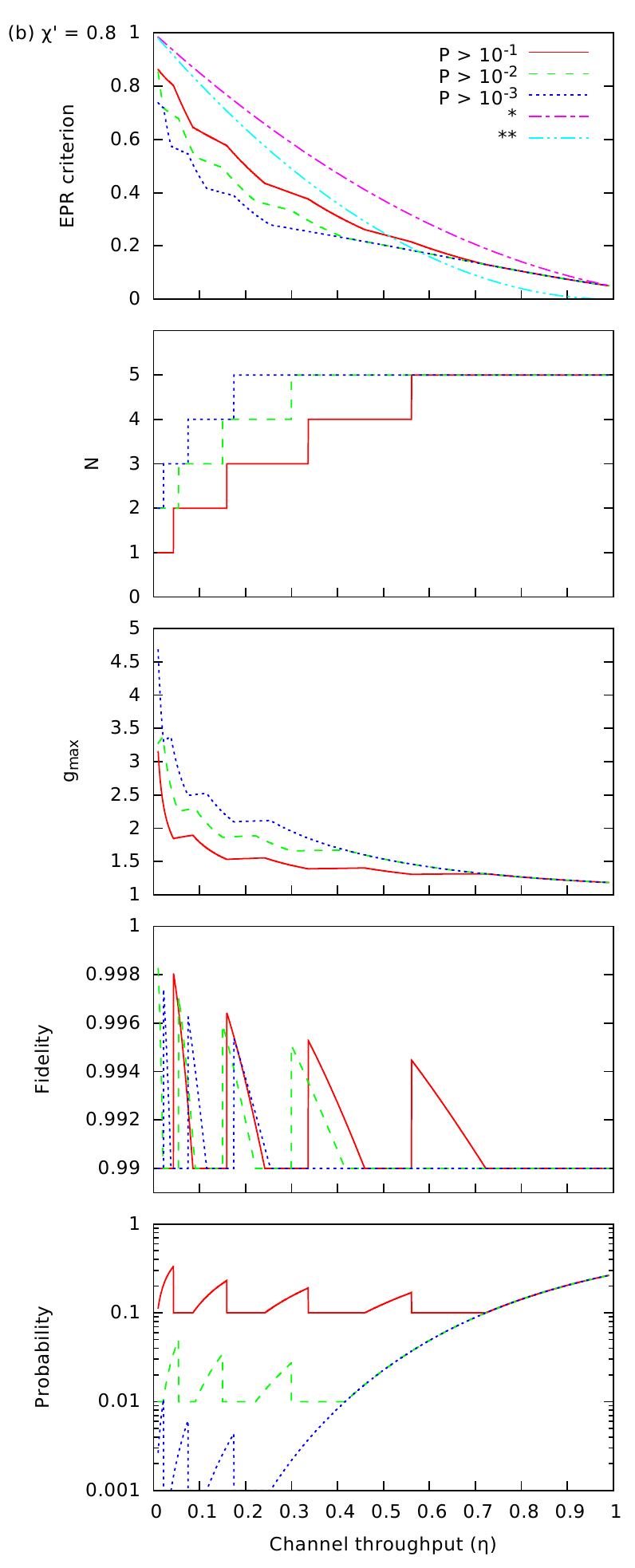}
  \caption{Plots of the lowest possible EPR criterion, $N$, gain $g$, lower
	  bound of fidelity and probability of success which achieve the of
	  lowest possible EPR criterion, $\varepsilon_{EPR}$ such that
	  $\mathcal{F}_{min} > 0.99$, probability of success $P > 0.1$ (solid
	  line), $0.01$ (dashed line) and $0.001$ (dotted line) and output
	  squeezing of $\chi^\prime = 0.5$ shown in (a) and $\chi^\prime = 0.8$
	  shown in (b).  The x-axis of each plot is the channel throughput
	  $\eta$.  The plots showing the optimised EPR criterion have curves
	  showing the EPR criterion when no amplification is performed (*,
	  dash-dot line) and the EPR criterion with no amplification with an
	  infinitely squeezed source with the same loss (**, dash-dot-dot
          line).}
  \label{fig:BestEPRCond}
\end{figure*}

To consider both a probability and fidelity bound we consider a numerical
optimisation of the EPR criterion for an amplified EPR state which results a
particular output squeezing $\chi^\prime$ which has undergone one sided loss
$1-\eta$.  The optimisation we will consider here enforces a fidelity greater
than $0.99$ and the probability of success greater than either $0.1$, $0.01$
and $0.001$.  Because of the monotonic nature of the fidelity and probability
under such conditions, we find that this optimisation always occurs on the
boundary of either the probability constraint or the fidelity constraint.
Figure~\ref{fig:BestEPRCond} shows the results of this optimisation when
$\chi^\prime = 0.5$ and $0.8$ as a function of loss.

The results of this optimisation are best understood by starting at the case
where $\eta=1$.  For this case we want to find if we are at the
boundary of the fidelity or probability constraints whilst ensuring that both
constraints are satisfied.  Also, the largest possible gain which achieves the
fidelity constraint will occur at the lowest value for $N$.  Therefore we seek
the gain and lowest $N$ such that our fidelity and probability constraints are
satisfied.  As the loss is increased, less signal is amplified and the fidelity
and probability increase.  Therefore a larger gain can be chosen which still
satisfies the constraints.  This continues until such point as the input signal
is weak enough so that the next lowest $N$ satisfies the constraints.  This
results in a discontinuous jump in the output.  Also, if the probability was
the saturated constraint, when $N$ is decremented this may change to the
fidelity constraint being the one that is saturated.  As loss is increased
further there will be a point where the saturation of these constraints will
swap.  This results in sharp corners appearing in the maximised curves for the
gain and EPR criterion.

The figures also show a comparison of this best EPR criterion to particular
situations not involving any amplification process.  The amplification process
always produces a lower EPR criterion when compared with doing no
amplification.  However, it is probably of more interest to compare the
situation to that of assuming the entanglement source could in principle
produce a maximally entangled EPR state (i.e. $\chi = 1$).  Because of the
loss, the EPR criterion for this limiting case is not zero.  Our amplification
model can succeed in producing a lower EPR criterion than that of the maximally
entanged  source. As shown in Figure~\ref{fig:BestEPRCond} this improvement
occurs in high loss situations.  The parameters for which this improvement
occurs will depend on the value of $\chi^\prime$ chosen.  But as shown in
figure~\ref{fig:BestEPRCond} the range of losses for which this occurs can
cover a significant range.

\section{Conclusion}

This paper has demonstrated a new model which could be used as a noiseless
phase insensitive linear amplifier.  We have presented a unitary for the
non-conditional evolution of a coupled harmonic oscillator system and a
heralding qubit. This evolution can then be used as a probabilistic amplifier
by measuring the heralding qubit after the unitary evolution.  The evolution is
not that of a linear optical transformation, but does achieve the highest
theoretically possible probability of success.  The action of our noiseless
amplification model on a coherent state and an EPR state was computed.  For an
EPR state undergoing one sided loss, we found that for sufficiently high loss
it is possible for the amplifier to achieve an EPR criterion lower than that
possible using an unamplified infinite squeezed source passing through the same
loss.  By choosing our parameters such that we target a particular level of
two-mode squeezing when the amplification succeeds, we have shown that, for the
case of single sided loss, the fidelity of the amplification has a lower bound.
This model and the results we have computed here may be used as a guide to
future experiments which wish to operate near the optimal probability of
success.

\section{Acknowledgements}

This research was conducted by the Australian Research Council Centre of
Excellence for Quantum Computation and Communication Technology (Project
number CE110001027).


\begin{thebibliography}{99}
\bibitem{betafunc} We have used the definition of the beta function as
	$B_z(a,b)=\int_0^z u^{a-1} (1-u)^{b-1} \mathrm{d}u$ and the
	regularlised beta function is then $I_z(a,b) = B_z(a,b)/B_1(a,b)$.
\end{thebibliography}
\end{document}